\documentclass{article}
\usepackage{setspace}
\usepackage{graphicx}
\usepackage{dcolumn}
\usepackage{bm}
\usepackage{xcolor} 
\usepackage{authblk}
\usepackage[mathlines]{lineno}
\usepackage[
    left=2cm,   
    right=2cm,  
    top=2cm,    
    bottom=2cm, 
]{geometry}
\usepackage{amsmath}

\begin{document}


\title{Strain-rate, temperature and size effects on the mechanical behavior of fiber bundles}

\author{Jérôme Weiss}
\affil{IsTerre, CNRS/Université Grenoble Alpes, Grenoble,France}

\maketitle

\begin{abstract}
The mechanical characteristics of fibers (of various materials), as well as of fiber bundles, are of primary importance for the design and the mechanical behavior of textiles, or of fibrous and composite materials. These characteristics are classically determined from strain-rate controlled tensile testing, generally assuming a negligible role of thermal activation on damage and fracturing processes. Under this assumption, the distribution of individual fiber strengths can be deduced from a downscaling of the macroscopic mechanical behavior at the bundle scale. 
There are however many experimental evidences of strain-rate and temperature effects on the mechanical behavior of individual fibers or bundles, which can also creep under constant applied load. This indicates a strong role of thermal activation on these processes. Here, these effects are analyzed from a fiber-bundle model with equal-load-sharing, in which thermal activation of fiber breakings is introduced from a kinetic Monte-Carlo algorithm adapted for time-varying stresses. This allows to rationalize these rate or temperature effects, such as a decrease of bundle strength, strain at peak stress, and apparent Young's modulus with decreasing strain-rate and/or increasing temperature. This also shows that the classical downscaling procedure used to estimate the distribution of individual fiber strengths from the mechanical behavior at the bundle scale should be considered with caution. If mechanical testing of the bundle is performed under conditions favoring the role of thermal activation (e.g. low applied strain-rate), this procedure can strongly underestimate the intrinsic (athermal) Weibull's parameters of the fiber strengths distribution. The same model is used as well to explore size (number of fibers) effects on bundle mechanical response. 

\end{abstract}



\section{\label{sec:Intro}Introduction}

The mechanical behavior of individual fibers, and of fiber bundles, is of primary importance for the design and the mechanical properties of textiles, as well as fibrous or composite materials, especially when the matrix is soft compared to the fibers, while fiber reinforced composites are of strong interest in various field of engineering, in particular owing to their high strength-to-weight ratio and fatigue resistance (e.g.\cite{prashanth2017fiber}).
When the athermal tensile strength of individual fibers has been measured \cite{wagner1984study,chi1983experimental}, the associated statistical variability has been generally considered to follow Weibull statistics \cite{weibull1939statistical}, therefore arguing for a weakest-link framework. In this case, the cumulative probability of fiber failure under an applied stress $\sigma$, among a population of fibers, reads:
\begin{equation}
    P_F(\sigma,L)=1-\text{ exp} \left [-\left(\frac{L}{L_0}\right)^d \left (\frac{\sigma}{\sigma_u}\right)^m \right]
    \label{eq:weibull}
\end{equation}
, where $L$ is the system size, $d$ the topological dimension, $m$ the Weibull's modulus, and $L_0$ and $\sigma_u$ normalizing constants. In case of individual fibers with a very small diameter/length ratio, $L$ reduces to the fiber length $l$, and $d$=1. To check that a weakest-link approach is relevant, the scale-dependence of $P_F$, as described in eq. (\ref{eq:weibull}), should be checked from a comparison of datasets obtained from different system sizes $L$ \cite{weiss2014finite}, which is rarely done.

Some authors also considered that Weibull's statistics can describe the strength variability of fiber bundles \cite{bai2020quasi,naito2023effect}. In this case, considering a fixed fiber length $l$, the system size $L$ in eq.(\ref{eq:weibull}) scales as the number of fibers in the bundle, $L\sim N_0$, and $d=1$. 

To determine the strength variability of individual fibers can be challenging, tedious and time-consuming. Consequently, it has been proposed, assuming a Weibull's strength distribution for fibers, that the parameters of this distribution ($m,\sigma_u,..$) can be obtained from the (macroscopic)stress-strain curves of fiber bundles \cite{chi1984determination,callaway2017strengths}, following several assumptions for the upscaling from the fiber individual scale to that of the bundle: (i) a monotonic strain-controlled tensile loading mode is applied to the bundle, (ii) all fibers are loaded uniformly and have the same Young's modulus $Y$ and the same cross-section $A$, (iii) the mechanical response of the bundle results from the successive and deterministic ruptures of fibers, from the weakest to the strongest one, i.e. through a purely athermal process. From this, a deterministic expression for the engineering, or macroscopic stress($\sigma_M$)-strain($\varepsilon$) relationship of the bundle can be obtained \cite{callaway2017strengths}:
\begin{equation}
    \sigma_M(\varepsilon)=Y\varepsilon \text{ exp} \left [-\frac{l}{l_0} \left (\frac{Y\varepsilon}{\sigma_u}\right)^m \right]
    \label{eq:SScurve}
\end{equation}
, where $l$ is the fibers length and $l_0=L_0$. The (athermal) bundle strength $\sigma_{f,ath}$ as well as the corresponding strain at peak stress $\varepsilon_{f,ath}$ can then be obtained from eq.(\ref{eq:SScurve}) with the condition $d\sigma/d\varepsilon=0$:
\begin{equation}
    \sigma_{f,ath}=\sigma_u \left(\frac{lme}{l_0}\right)^{-1/m} \text{ and  } \varepsilon_{f,ath}=\frac{\sigma_u}{Y} \left(\frac{lm}{l_0}\right)^{-1/m}
    \label{eq:failureStrength}
\end{equation}
, where $e\simeq 2.718$ is the base of the natural logarithm. This gives a size (i.e. fiber length $l$) effect on strength $\sigma_{f,ath}$ as well as on strain at peak stress $\varepsilon_{f,ath}$, but no dependence on the number of fibers $N_0$ within the bundle.
Then, the parameters of the Weibull's distribution of fiber strengths can be in principle obtained from tensile tests performed on bundles and inverting eq.(\ref{eq:SScurve}) or eq.(\ref{eq:failureStrength}) \cite{callaway2017strengths}.

However, in all the works mentioned above, the rupture of each fiber is, explicitly or implicitly, assumed to occur deterministically once the nominal fiber strength is reached. This precludes strain-rate or temperature effects. This is in contradiction with many experimental evidences obtained for various materials. The bundle strength $\sigma_{f}$, the strain at peak stress $\varepsilon_f$, as well as the apparent Young's modulus of the bundle $Y_b$ generally decrease with decreasing applied strain-rate, both in the quasi-static ($\dot{\varepsilon}<1 s^{-1}$) and dynamic ($\dot{\varepsilon}\gg1 s^{-1}$) regimes (e.g. \cite{wang1998effects,wang2024effects} for Aramid fibers, \cite{bai2020quasi,zhao2023effects} for PET, \cite{zhao2023effects} for PEN, \cite{huang2003experimental} for SiC,...), with an exception for carbon fiber bundles that seem essentially strain-rate insensitive\cite{zhou2010tensile}. In what follows, we will focus on strain-rate and temperature effects in the quasi-static regime, leaving aside dynamical effects.
Similar trends have been reported as well for the mechanical properties of single fibers (e.g. \cite{sun2012modified,wang2020tensile}). In addition, when subjected to a constant load, below their athermal strength, individual fibers, bundles, or fibrous materials creep, i.e. slowly deform under a constant applied load, and eventually break after some time $t_f$\cite{coleman1957time,wagner1986lifetime,rosti2010fluctuations,koivisto2016predicting}, with increasing temperatures increasing strain-rates and decreasing lifetimes \cite{busse1942fatigue,wu1988temperature}. All of this argues for a key role of thermal activation in the deformation, damage and rupture processes of fibers, bundles or composite materials.

Starting with Coleman \cite{coleman1956time,coleman1957time,coleman1958statistics}, fiber-bundles-models (FBM) have been the privileged tool to study the time-dependent rupture of bundles. Coleman introduced a so-called "theory of breaking kinetics" based on several assumptions: (i) all unbroken fibers share equal load, corresponding to an equal load sharing (ELS) mode suited to describe composite materials with a soft matrix relatively to the fiber stiffness, (ii) the bundle is homogeneous, i.e. there is no variability in the fiber strength, and (iii) the time-dependent rupture of fibers is controlled by an empirical "breaking rule", identical for each fiber and that gives the probability for a fiber to break before time $t$. Later, Phoenix and co-workers \cite{wagner1986lifetime,phoenix1983statistical} justified Coleman's theory of breaking kinetics by making a link with reaction rate theory \cite{eyring1935activated}. In reaction rate theory, the rate of breaking of a fiber under a stress $\sigma$ reads:
\begin{equation}
    r=\omega_0 \text{exp}\left(\frac{-U}{k_BT}\right)
    \label{eq:breakingrate}
\end{equation}
, where $U$ is an activation energy, $\omega_0$ an attempt frequency, $k_B$ the Boltzmann's constant and $T$ the temperature. The activation energy $U$ is linked to the stress gap between the athermal strength of the fiber and the stress applied on the fiber, $\Delta\sigma=\sigma_s-\sigma$. A linear relation, $U=V_a\Delta\sigma$ with $V_a$ an activation volume, corresponding to an exponential breaking rule in Coleman's framework, was proposed by Eyring \cite{eyring1935activated,eyring1936viscosity}, but $U\sim \Delta\sigma^{3/2}$ has been considered as well \cite{maloney2006energy}, while Phoenix and Tierney \cite{phoenix1983statistical} proposed $U\sim -\text{log}(\sigma/\sigma_s)$ corresponding to a power-law Coleman's breaking rule. 

As stressed above, within a bundle, the strength varies among fibers, i.e. disorder on $\sigma_s$, and consequently on the activation energy $U$ should be considered, even if the applied stress $\sigma$ is assumed the same for all fibers (ELS mode). Several authors considered the thermally-activated deformation and rupture of disordered bundles using FBM, either by adding a thermal white noise $\xi_i$ to the stress applied on each fiber $i$ \cite{roux2000thermally,ciliberto2001effect,yoshioka2008size}, or by using a more efficient kinetic Monte-Carlo algorithm \cite{bortz1975new} to activate fiber breakings \cite{weiss2023logarithmic,verano2024effect,verano2025creep,hiemer2025fiber}. 

However, all these works considered a creep loading mode, i.e. a constant applied load $\mathcal{L}$, and therefore a constant applied stress $\sigma=\frac{\mathcal{L}}{AN}$ on each fiber between successive fiber breakings, for a bundle of $N$ remaining fibers of constant cross-section $A$. In what follows, in order to analyze strain-rate effects on the mechanical behavior of fiber bundles, a thermally-activated FBM is instead considered  under a strain-rate controlled loading mode, i.e. an increasing stress $\sigma(t)$ between successive fiber breakings.

Finally, only a few works considered the effect of the number of fibers in the bundle on the bundle strength, while keeping the fibers length $l$ fixed \cite{zhao2023effects,naito2023effect}. Overall, the mean strength $\langle\sigma_f\rangle$ as well as its variability (standard deviation, $\delta\sigma_f$) among different tests, are observed to decrease with increasing size $N_0$. This decrease is moderate, by a factor from $\sim0.8$ to $\sim0.6$ for $N$ varying from 10 to $\sim10^4$ for different types of carbon fibers, while the decrease is more pronounced for $\delta\sigma_f$ over the same range of sizes (decreasing factor between 0.65 and 0.2) \cite{naito2023effect}. We first note that eq. (\ref{eq:failureStrength}) cannot account for such size effects. As later discussed in section \ref{subsec:size effects}, this comes from the fact that the simple scenario giving this expression ignores the critical character of bundle failure \cite{pradhan2010failure}, and associated finite-size effects.
On the reverse, an alternative explanation for these size effects could be a weakest-'link' (here, weakest-fiber) framework \cite{bai2020quasi,naito2023effect}, for which the strength of the weakest fiber would dictate the bundle strength. In this case, from eq. (\ref{eq:weibull}) with $N_0=L$ and $d=1$, one would expect the cumulative probability distribution of bundle failure stresses, $P_F(\sigma,N_0)$, to be shifted towards smaller stresses with increasing size $N_0$, but, by construction, the Weibull's modulus $m$ to be size-independent. This last point is in contradiction with Naito \cite{naito2023effect}, who reported an increasing $m$ with increasing size $N_0$, therefore raising doubt on such a weakest-fiber hypothesis. This point will be discussed further in section \ref{subsec:size effects}.
\\ 

\section{\label{sec:model}The model}

A bundle of (initially) $N_0$ parallel elastic fibers of unit length $l=1$ and cross section $A=1$ is loaded in an ELS mode under an imposed constant \textit{strain-rate} $\dot{\varepsilon}$ until final failure. This fundamentally differs from the classical stress-control mode considered in most FBM studies. This strain-rate control mode implies the introduction of time, done from the integration of thermal activation in the model (see below).
The elastic stiffness is the same for all fibers, $Y=1$. It is in principle possible to introduce some variability on $Y$, which would induce a variability of the stressing rate $\dot{\sigma}$ among fibers, but, for a sake of simplicity, this is not considered here. The material disorder is introduced only from a quenched distribution of athermal fiber strengths $\sigma_{s,i}$, with $i=1..N_0$. In what follows, two types of distribution are considered: (i) a uniform distribution between 0 and 1, as classically considered in FBM studies (e.g. \cite{alava2006statistical,pradhan2010failure}), (ii) or, following observations on individual fibers (see above), Weibull distributions of shape parameter $m$ and scale parameter $\sigma_u=0.5$ (in order to have a mean $\langle\sigma_s\rangle=\sigma_u\Gamma(1+1/m)$, where $\Gamma$ is the Gamma function, similar to that of the uniform distribution).

Calling $\mathcal{L}$ the applied load, the stress on the fibers is $\sigma=\mathcal{L}/N$ while the macroscopic(engineering) stress on the bundle is: 
\begin{equation}
\sigma_M=\mathcal{L}/N_0=\sigma\frac{N}{N_0}
\label{eq:macrostress}
\end{equation}
, with $N$ the number of unbroken fibers.
Under a strain-rate controlled mode, this expression shows that the evolution of the engineering stress with applied strain results from a competition between an increasing local stress $\sigma(t)$ and a decreasing $N(t)$, which implies a maximum value corresponding to the bundle strength $\sigma_f$. The fibers being elastic, one has $\sigma=Y\cdot\varepsilon$, and therefore $\dot{\sigma}=Y\cdot\dot{\varepsilon}$. The initial conditions ($t=0$) are $N=N_0$ and $\varepsilon=\mathcal{L}=\sigma=\sigma_M=0$.

We consider that the deformation of the bundle results from a succession of thermally activated individual fiber breakings separated by waiting times $\Delta t_j$, therefore introducing time in the model. Under such a strain-rate controlled mode, the strain and therefore the local stress $\sigma$ do not change after a fiber breaking, and therefore athermal avalanches of breakings are not possible. Thermal activation of fiber breaking is modeled using a dynamical Monte-Carlo algorithm adapted for time-dependent stresses, and so transition rates \cite{prados1997dynamical}. This consists in two steps:

(Step 1). The distribution of fiber strengths implies at each transition a distribution of $N$ stress gaps $\Delta\sigma_i=\sigma_{s,i}-\sigma$, leading to transition rates:
\begin{equation}
r_i=\omega_0\text{exp}\left(\frac{-V_a\Delta\sigma_i}{k_BT}\right)=\omega_0\text{exp}\left(\frac{-\Delta\sigma_i}{\theta}\right)
\label{eq:rates}
\end{equation}
, where we assume an Eyring-like linear scaling for the activation energy and we call later on 'temperature' the stress scale $\theta=\frac{k_BT}{V_a}$, while $t_0=1/\omega_0$ defines the reference timescale.
The waiting time till the next transition, $\Delta t$, is given by:
\begin{equation}
\int_0^{\Delta t}\sum_{i=1}^N r_i(t')dt'=-\text{log}(u)
\label{eq:waitingtimegeneral}
\end{equation}
, where $u\in[0..1]$ is a randomly chosen number. In case of a creep loading mode, as explored by \cite{weiss2023logarithmic,verano2025creep,hiemer2025fiber}, the stress gaps $\Delta\sigma_i$ are unchanged between two successive transitions, and so, by integration of eq.(\ref{eq:waitingtimegeneral}), $\Delta t=\frac{-\text{log}(u)}{\omega_0Z}$, where $Z=\sum_{i=1}^N\text{exp}(\frac{-\Delta\sigma_i}{\theta})$ is the partition function of the system at the previous transition \cite{verano2024effect}.

In the case of a strain-controlled mode, as considered here, the stress gaps between two successive transitions decrease at a rate $\dot{\varepsilon}\cdot Y$, constant for all fibers and all transitions, i.e. $\Delta\sigma_i(t)=\Delta\sigma_{i,prev}-\dot{\varepsilon}\cdot Y\cdot t$, where $\Delta\sigma_{i,prev}$ is the stress gap of fiber $i$ after the previous transition. Introducing this into eqs. (\ref{eq:rates}) and (\ref{eq:waitingtimegeneral}), and after some algebra, the following expression for the waiting time is obtained: 

\begin{equation}
\Delta t=\frac{\theta}{\dot{\varepsilon}\cdot Y}\text{log}\left(1-\frac{\dot{\varepsilon}\cdot Y\text{log(u)}}{\theta \omega_0 Z}\right)
\label{eq:waitingtime}
\end{equation}

This waiting time is compared to the time $\Delta t_{ath}$ needed, under the imposed strain-rate, to fill entirely the smallest stress gap $\Delta\sigma_{min}$, i.e. to break athermally the weakest fiber. If $\Delta t_{ath}<\Delta t$, this fiber is selected for breaking, and $\Delta t$ is set to $\Delta t_{ath}$.
If not, the algorithm goes to Step 2. In practice, such athermal breaking becomes significant only at very large strain-rates, see section \ref{subsec:strain-rate effects}.

(Step 2). The fiber that breaks at this transition is randomly chosen, according to the probabilities given by eq.(\ref{eq:rates}). After this, the number of unbroken fibers is decreased, $N\rightarrow N-1$, the time advances $t\rightarrow t+\Delta t$ and consequently the strain $\varepsilon\rightarrow\varepsilon+\Delta t\cdot\dot{\varepsilon}$ and the local stress $\sigma\rightarrow\sigma+\Delta t\cdot Y\cdot\dot{\varepsilon}$, while the engineering stress $\sigma_M$ is adjusted from eq. (\ref{eq:macrostress}).

In this framework, when the temperature vanishes ($\theta\rightarrow0+$), the dynamics is expected to become extremal, that is, the fiber that breaks is always the one with the smallest stress gap, like under very large strain-rates. Consequently, much like what is often assumed \cite{chi1984determination,callaway2017strengths}, the bundle deformation results from a deterministic succession of fiber ruptures, from the weakest to the strongest one. Considering first a uniform distribution of fiber strengths, the cumulative probability of failure is simply $P_F(\sigma)=\sigma$. Therefore, at an applied strain $\varepsilon$, the number of surviving fibers is $N=N_0(1-P_F(\sigma))=N_0(1-\sigma)$, and so, from eq. (\ref{eq:macrostress}), the engineering stress-strain relationship is given by:
\begin{equation}
\sigma_M(\varepsilon)=\varepsilon\cdot Y(1-\varepsilon\cdot Y)
\label{eq:SScurveUniform}
\end{equation}
, which is symmetrical with respect to the peak stress $\sigma_{f,th}=0.25$ reached at strain $\varepsilon_{f,th}=0.5$, without dependence on the bundle size $N_0$. This peak stress is the same that the one reached asymptotically for a classical ELS FBM with a very large $N_0\rightarrow+\infty$, but without thermal activation, under a \textit{stress}-controlled mode \cite{pradhan2010failure,roy2013scaling}.
For a Weibull distribution, a similar reasoning gives eq. (\ref{eq:SScurve}), with $l=l_0=1$ in our case \cite{chi1984determination}. These two expressions (eqs. (\ref{eq:SScurveUniform}) or (\ref{eq:SScurve})) will represent our two strain-stress relationships of 'reference' for an athermal case, i.e. when the role of thermal activation vanishes.

In what follows, all the results are given in terms of dimensionless variables, with $\sigma_{ref}=1$ the reference stress corresponding to the maximum possible fiber strength for the uniform distribution, $\varepsilon_{ref}=1$ the reference strain corresponding to the maximum strain possible for a uniform fiber strength distribution at vanishing temperature, $t_{ref}=t_0$ the reference time, and consequently $\dot{\varepsilon}_{ref}=\varepsilon_{ref}/t_{ref}=1/t_0=\omega_0$ the reference strain-rate. Consequently, this model is physically meaningful only for $\dot{\varepsilon}\ll1$. The results discussed in sections \ref{subsec:strain-rate effects} and \ref{subsec:temperature effects} were obtained for a large number of fibers in the bundle, $N_0=10^5$, and all the  SS-curves shown averaged over 50 realizations of disorder. For the analysis of size effects in section \ref{subsec:size effects}, $N_0$ varied from 10 to $10^5$.

\section{\label{sec:results}Results}

\begin{figure}
\includegraphics[width=\linewidth]{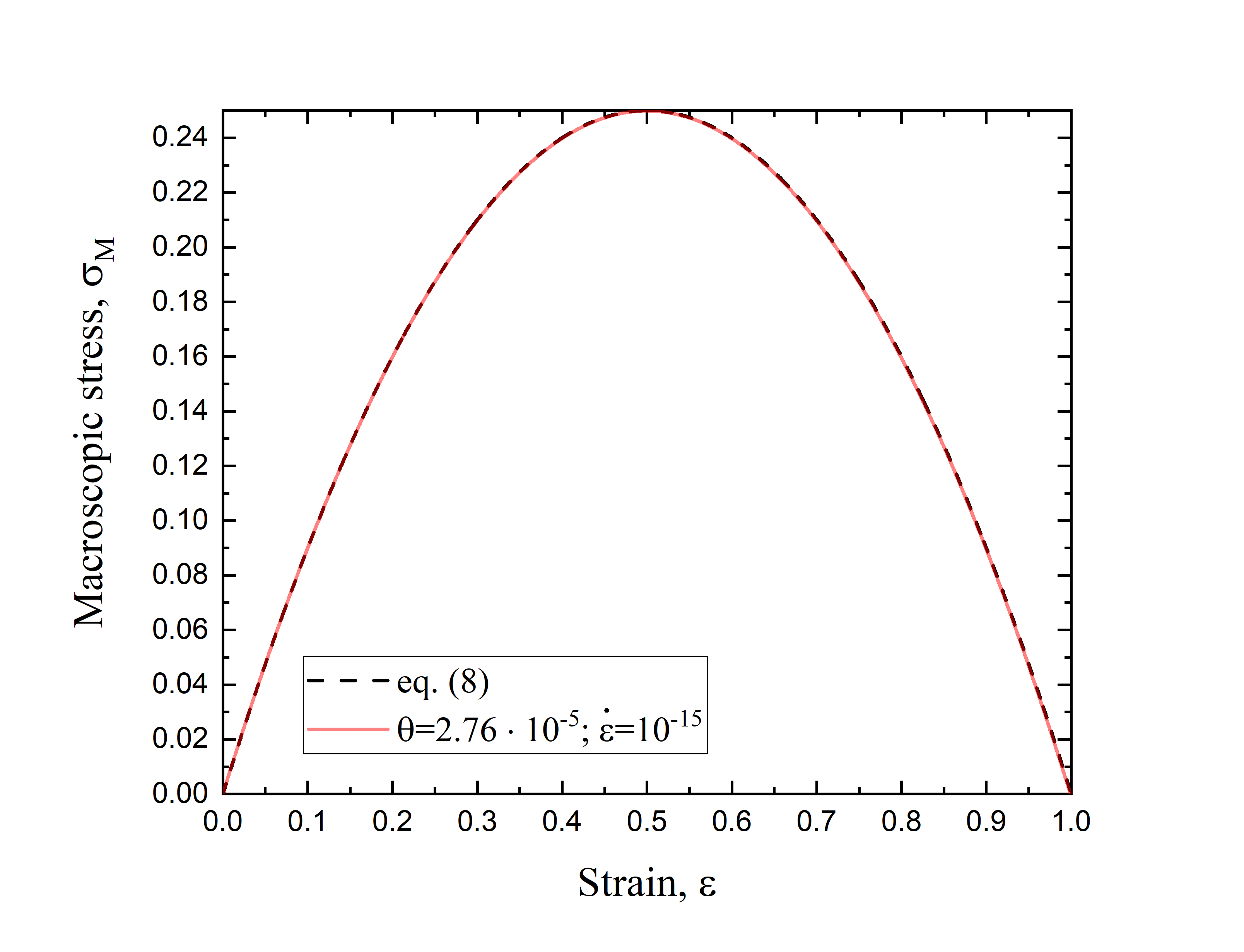}
\caption{\label{fig:Athermal_uniform} Macroscopic stress-strain curves of bundles, averaged over 50 realizations of disorder, for an intermediate strain-rate of $\dot{\varepsilon}=10^{-15}$, a vanishing temperature $\theta=2.76\times10^{-5}$ and a uniform distribution of individual fiber strengths. This is compared to expression (\ref{eq:SScurveUniform}) corresponding to athermal dynamics.}
\end{figure}

As a first test of the model, Figure \ref{fig:Athermal_uniform} shows a comparison of Strain-Stress (SS) curves obtained for an intermediate strain-rate of $\dot{\varepsilon}=10^{-15}$, a vanishing temperature $\theta=2.76\times10^{-5}$ and a uniform distribution of individual fiber strengths, with the expression (\ref{eq:SScurveUniform}) obtained by ignoring thermal activation. The numerical results and this theoretical prediction are undistinguishable, confirming the extremal character of the dynamics when $\theta\rightarrow0$. Note that, with these conditions of temperature and applied strain-rate, all individual fiber ruptures are thermally activated, but the extremal dynamics leads to a mechanical response identical to what is predicted assuming a purely athermal process. 

\subsection{\label{subsec:strain-rate effects}Strain-rate effects}

\begin{figure}
\includegraphics[height=0.8\textheight, keepaspectratio]{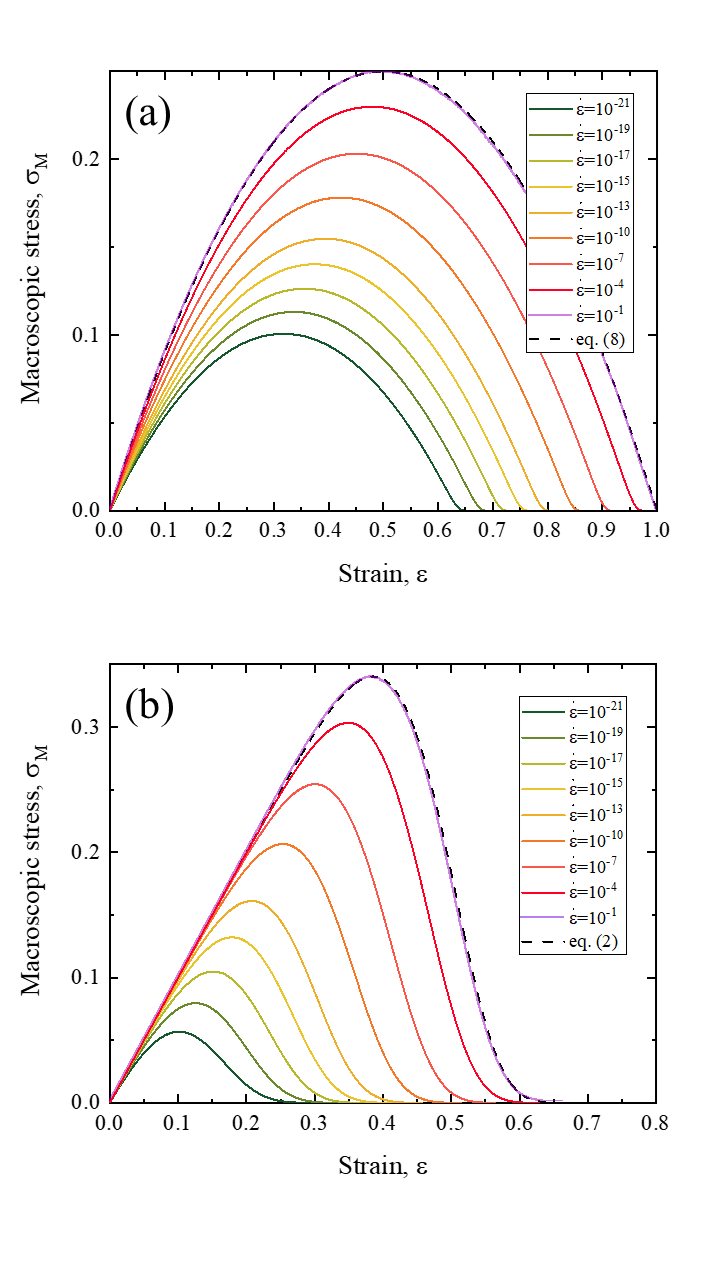}
\caption{\label{fig:Strain_rate_effects} Macroscopic stress-strain curves of bundles, averaged over 50 realizations of disorder, for an intermediate  temperature $\theta=8.28\times10^{-3}$, varying strain-rates, and (a) a uniform or (b) a Weibull's distribution ($m=8$) of individual fiber strengths. This is respectively compared to (a) expression (\ref{eq:SScurveUniform}) or (b) expression (\ref{eq:SScurve}) corresponding to athermal dynamics.}
\end{figure}

\begin{figure}
\includegraphics[height=0.8\textheight, keepaspectratio]{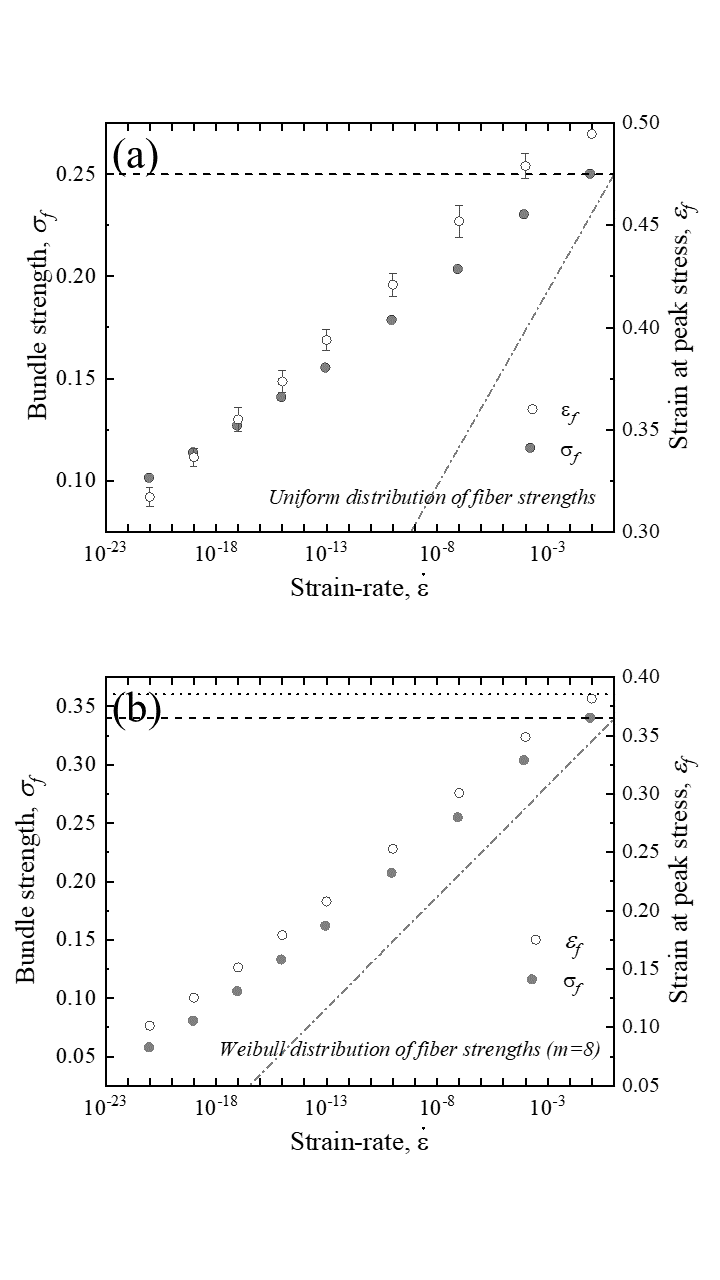}
\caption{\label{fig:Strain_rate_effects_2} Evolution of the average bundle strength $\langle\sigma_f\rangle$ (closed symbols), as well as of the average strain at peak stress $\langle\varepsilon_f\rangle$ (open symbols), with increasing strain-rate for (a) a uniform and (b) a Weibull ($m=8$) distribution of fiber strengths, and a temperature $\theta=8.28\times10^{-3}$. The diagonal dashed-dotted lines represent the prediction of eq.(\ref{eq:expectedStrength2}) for the strength. The dashed horizontal line indicates the athermal value of the bundle strength, and the dotted horizontal line that of the strain at peak stress, both recovered at very large strain-rates. The error bars represent the strength variability among 50 realizations of disorder. When this variability is smaller than the symbol size, it is not represented.}
\end{figure}

Figure \ref{fig:Strain_rate_effects} shows SS-curves obtained for different strain-rates, for a uniform (a) or a Weibull ($m=8$) (b) distribution of fiber strengths and an intermediate temperature $\theta=8.28\times10^{-3}$. The average bundle strength $\langle\sigma_f\rangle$ as well as the average strain at peak stress $\langle\varepsilon_f\rangle$ increase with increasing applied strain-rate, in agreement with observations \cite{bai2020quasi,wang1998effects,wang2024effects,zhao2023effects,huang2003experimental}, and the SS-curve converges towards the prediction for an athermal rupture process (eq. (\ref{eq:SScurveUniform}) for a uniform distribution of fiber strengths, eq. (\ref{eq:SScurve}) for a Weibull distribution) at very large strain-rates. Actually, the fraction of athermal fiber breaking is zero for $\dot{\varepsilon}\leq10^{-4}$ and switches to almost 1 for $\dot{\varepsilon}\geq10^{-1}$ in both cases. The growth of $\langle\sigma_f\rangle$ as well as of $\langle\varepsilon_f\rangle$ with $\dot{\varepsilon}$ is logarithmic, see Figure \ref{fig:Strain_rate_effects_2}. 

One can understand this decrease of the bundle strength with decreasing applied strain-rate from a competition of timescales. That is, in one hand, the time $\Delta t_{ath}$ needed to fill, athermally, a stress gap $\Delta \sigma=\sigma_{f,ath}-\sigma$ between a stress $\sigma$ and the athermal strength $\sigma_{f,ath}$, which is simply $\Delta t_{ath}=\frac{\Delta\sigma}{\dot{\varepsilon}Y}$ and, on the other hand, a typical activation timescale that reads $\Delta t=t_0\text{exp}\left(\frac{\Delta \sigma}{\theta}\right)$. A decreasing $\dot{\varepsilon}$ means that it takes a longer time to reach the athermal rupture of fibers, giving more time for thermally activation to take place and consequently to trigger a rupture at a smaller stress. In eq. (\ref{eq:waitingtime}), the activation time actually depends on strain-rate. However, the term $\frac{\dot{\varepsilon}\cdot Y}{\theta \omega_0 Z}$ is generally small compared to 1, except at vanishing temperature, and therefore $\Delta t\simeq\frac{-\text{log}(u)}{\omega_0 Z}$, which is strain-rate independent. This amounts to neglect the decrease of the stress gap between two successive transitions, and so the expression of $\Delta t$ under a creep loading is recovered. Then, writing $\Delta t_{ath}=\frac{\Delta\sigma}{\dot{\varepsilon}Y}=\Delta t=t_0\cdot\text{exp}\left(\frac{\Delta \sigma}{\theta}\right)$, one gets:

\begin{equation}
\frac{\Delta\sigma}{\theta}-\text{log}(\Delta\sigma)=-\text{log}(\dot{\varepsilon}\cdot Y\cdot t_0)
\label{eq:expectedStrength}
\end{equation}
, or:

\begin{equation}
\sigma=\sigma_{f,ath}-\theta(\text{log}(\Delta\sigma)-\text{log}(\dot{\varepsilon}\cdot Y\cdot t_0))
\label{eq:expectedStrength2}
\end{equation}

As $\text{log}(\Delta \sigma)$ is generally much smaller than $\Delta\sigma$ (for a temperature $\theta=8.28\times10^{-3}$, $\text{log}(\Delta \sigma)=\Delta\sigma$ for $\Delta\sigma\simeq0.029$), the above equations suggests an increase of the bundle strength with the logarithm of the strain-rate, in qualitative agreement with numerical results (see Figure \ref{fig:Strain_rate_effects_2}) and consistent with available observations. This is quite noticeable, owing to the simplifications considered here, in particular the fact that to obtain an activation time while elastically increasing the stress (eq. (\ref{eq:waitingtime})), an integration over time of the activation rates $r_i$ has to be considered, which is ignored in the rough calculation above. This agreement is however only qualitative, as the slope of $\text{log}(\dot\varepsilon)$ \textit{vs} $\sigma$ observed for numerical results is much smaller than $\theta$.
Eq. (\ref{eq:expectedStrength2}) predicts also a roughly linear decrease of the bundle strength with temperature, which will be discussed below in section \ref{subsec:temperature effects}. 

In experiments, a decrease of the apparent Young's modulus of the bundle, $Y_b$, with decreasing strain-rate is generally reported \cite{wang2024effects,wang1998effects,bai2020quasi,huang2003experimental}. An estimation of $Y_b$ is obtained from the slope of the experimental SS-curves at the early stage of deformation. In the model described above, the initial stiffness of the bundle, up to the first fiber breaking, is by construction independent of the applied strain-rate and equal to the fiber stiffness $Y$. However, it is clear from figure \ref{fig:Strain_rate_effects} that the smaller the strain-rate, the softer is the macroscopic behavior with an earlier departure from linear elasticity, which can be interpreted as a decreasing apparent Young's modulus in agreement with observations. 

\subsection{\label{subsec:temperature effects}Temperature effects}

\begin{figure}
\includegraphics[width=\linewidth]{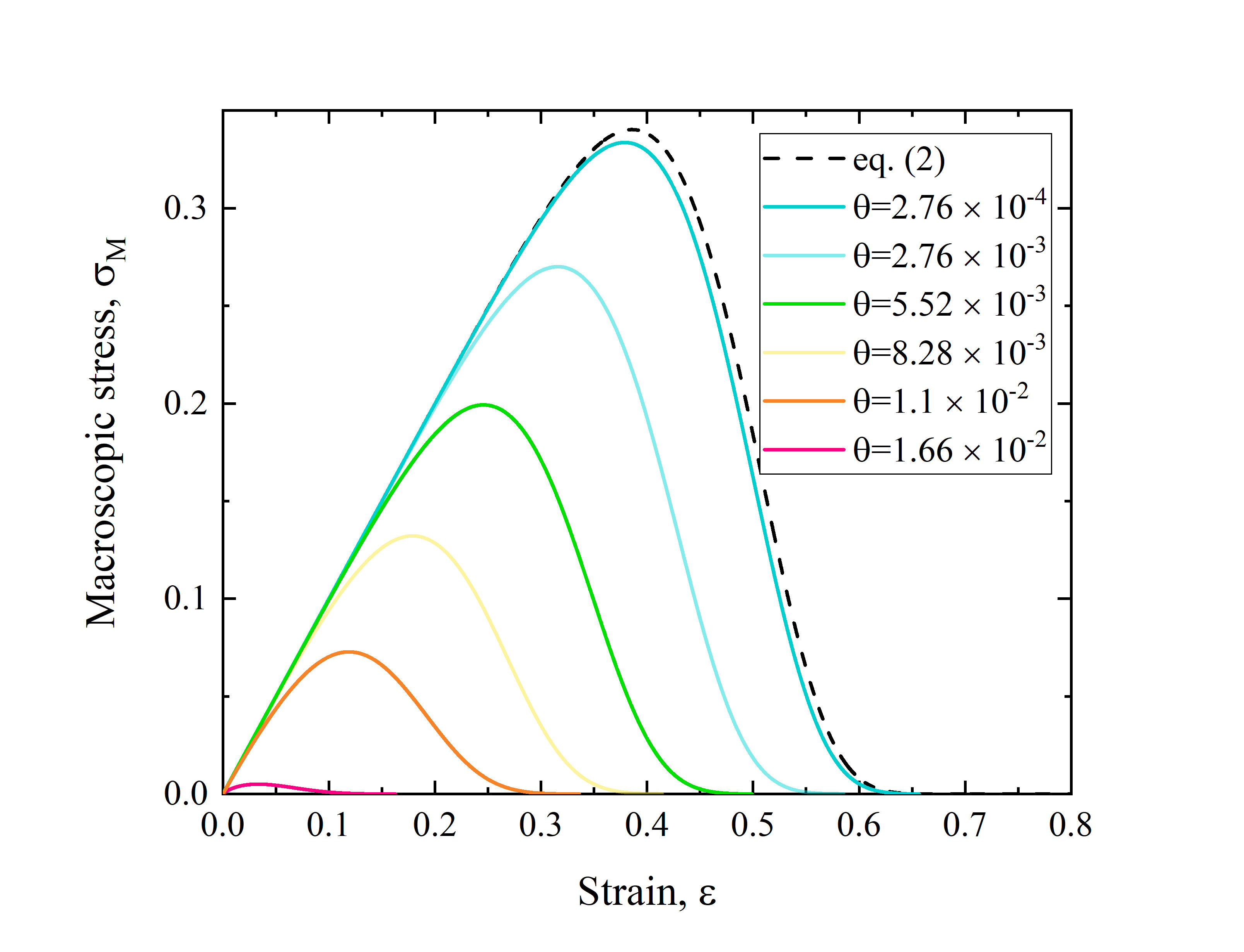}
\caption{\label{fig:temperature_effects} Macroscopic stress-strain curves of bundles for an intermediate  strain-rate $\dot{\varepsilon}=10^{-15}$, varying temperatures, and a Weibull's distribution of individual fiber strengths ($m=8$). This is compared to expression (\ref{eq:SScurve}) corresponding to athermal dynamics.}
\end{figure}

\begin{figure}
    \centering
    \includegraphics[width=1\linewidth]{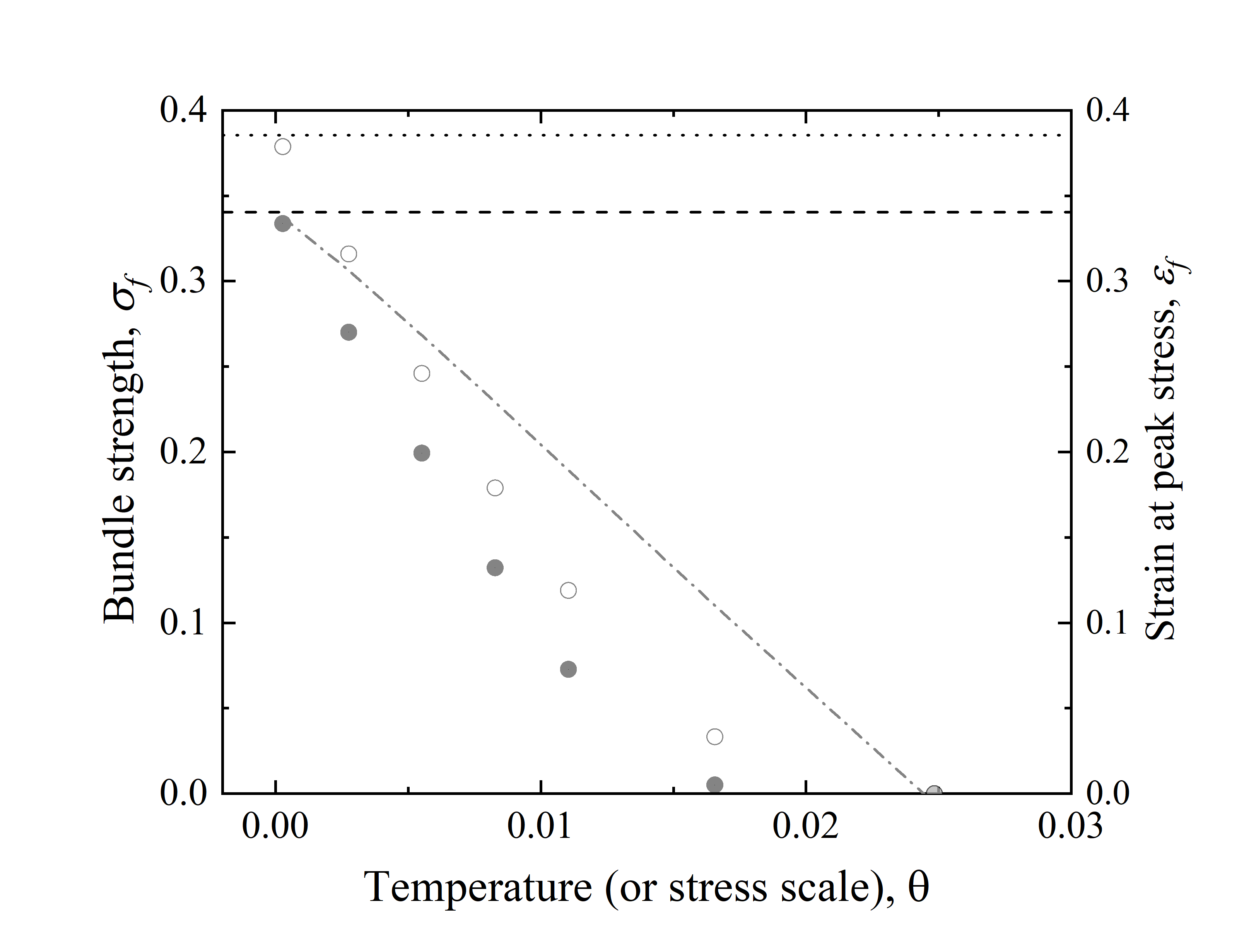}
    \caption{Evolution of the average bundle strength $\langle\sigma_f\rangle$ (closed symbols), as well as of the average strain at peak stress $\langle\varepsilon_f\rangle$ (open symbols), with increasing temperature $\theta$ for a Weibull ($m=8$) distribution of fiber strengths, and a strain-rate $\dot{\varepsilon}=10^{-15}$. The diagonal dashed-dotted line represents the prediction of eq.(\ref{eq:expectedStrength2}) for the strength. The dashed horizontal line indicates the athermal value of the bundle strength, and the dotted horizontal line that of the strain at peak stress, both recovered at very low temperature. The variability among 50 realizations of disorder was always smaller than the symbol size, and consequently not represented.}
    \label{fig:Strength_vs_T}
\end{figure}

Figure \ref{fig:temperature_effects} shows SS-curves obtained for a Weibull distribution ($m=8$), different temperatures, but a fixed strain-rate $\dot{\varepsilon}=10^{-15}$. Temperature effects for a uniform distribution are similar. The bundle strength as well as the strain at peak stress both decrease with increasing temperature, and almost vanish for $\theta>0.015$. On the reverse, for very low temperatures, the behavior converges towards the athermal expectation (eq. (\ref{eq:SScurve})), as noted above for a uniform distribution of fiber strengths. At low to intermediate temperatures ($\theta\leq0.1$), the strength decreases roughly linearly with increasing temperature, as suggested by the simple reasoning developed above in section (\ref{subsec:strain-rate effects}), but the slope is larger than expected (Fig. \ref{fig:Strength_vs_T}). This indicates that the argument based on a competition of two timescales qualitatively captures the trends, but not quantitatively, much like for strain-rate effects. The nearly vanishing strength at very large temperatures can be interpreted as an almost spontaneous activation of fiber breakings from strong thermal fluctuations. 

\subsection{\label{subsec:size effects}Size effects on bundle strength}

Here we consider the possible effect of the initial number of fibers $N_0$ within the bundle on its strength, while the fibers length is fixed. This fundamentally differs from the effect of fiber length $l$ on the average strength of individual fibers, which can be related to the fibers Weibull's statistics, see e.g. \cite{andersons2002glass}.
We have mentioned above that, when reported, the effect of an increasing number $N_0$ of fibers within a bundle is to moderately decrease the average bundle strength $\langle\sigma_f\rangle$, and to a larger extent the associated variability $\delta\sigma_f$ \cite{naito2023effect,zhao2023effects}. Some authors proposed to describe the statistical strength variability of bundles from Weibull's statistics, hence assuming the relevance of a weakest-link framework \cite{bai2020quasi,naito2023effect}. In this case, one would expect eq. (\ref{eq:weibull}), with $L\sim N_0$ and $d=1$, to describe bundle strength statistics, and
 the mean strength $\langle\sigma_f\rangle$ as well as the associated standard deviation to decrease as $\langle\sigma_f(N_0)\rangle\sim\delta\sigma_f(N_0)\sim N_0^{-1/m}$ \cite{weiss2014finite}. As the shape parameters reported for experimental datasets range between about 5 and 20, this predicts a weak size effect, possibly compatible with observations \cite{naito2023effect,zhao2023effects}, but cannot explain a stronger size effect on variability $\delta\sigma_f$. However, this would predict a vanishing mean strength for $N_0\rightarrow\infty$, which is counterintuitive. In addition, as already stressed, such weakest-fiber framework would imply a $m$-value independent of $N_0$, in contradiction with reported results \cite{naito2023effect}, consequently raising doubt about the applicability of this framework for fiber bundles.

\begin{figure}
\includegraphics[width=\linewidth]{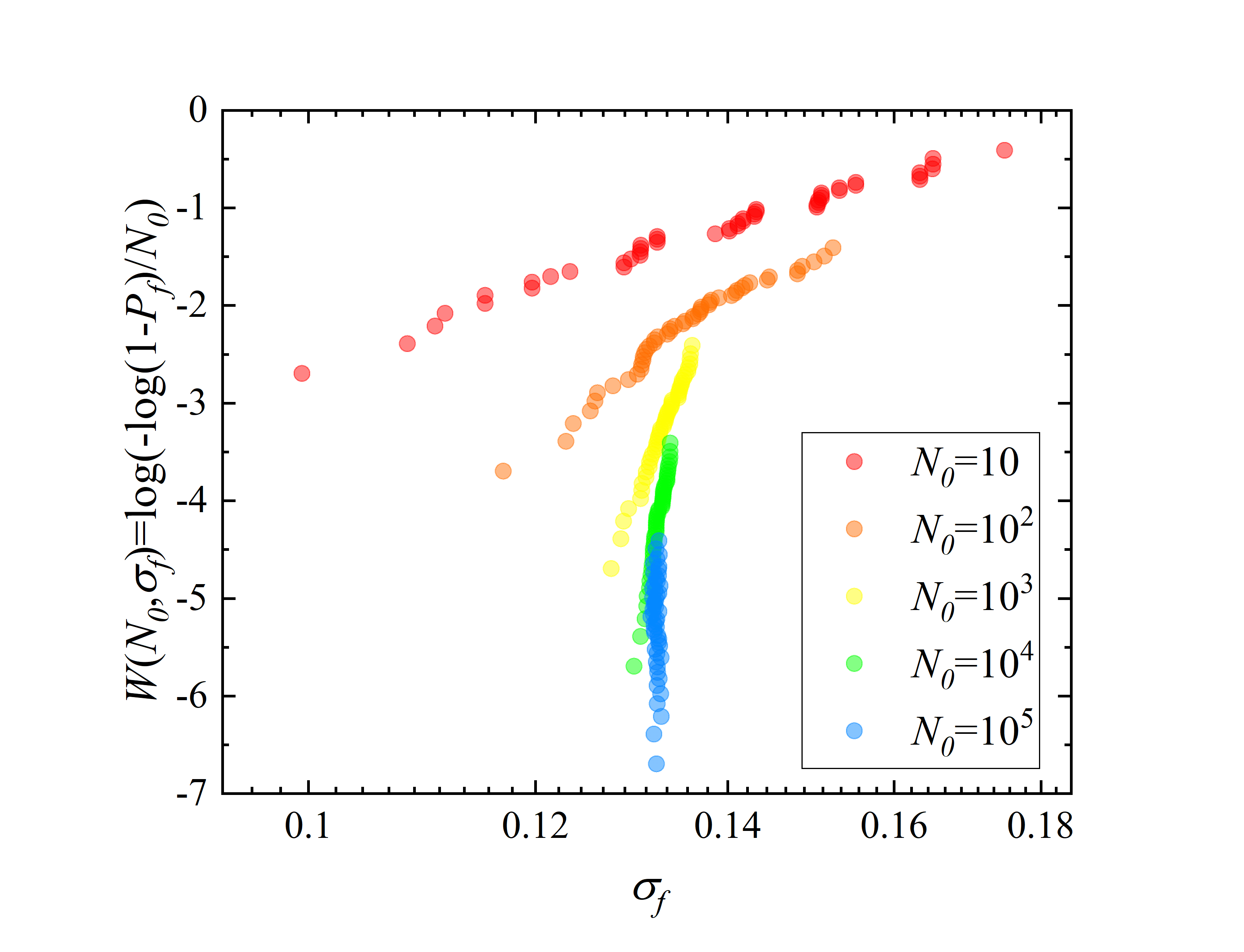}
\caption{\label{fig:Weibull_plots} Representation of 50 realizations of the model in a Weibull's plot, for a Weibull's distribution of fiber strengths ($m=8$), a strain-rate $\dot{\varepsilon}=10^{-15}$, a temperature $\theta=8.28\times10^{-3}$ and a varying initial number of fibers $N_0$. If Weibull's statistics would be relevant at the bundle scale, the data obtained for different sizes $N_0$ should collapse in such a representation \cite{weiss2014finite}.}
\end{figure}

On figure \ref{fig:Weibull_plots}, the quantity $W(N_0,\sigma_f)=\text{log}\left(-\text{log}\left(\frac{1-P_F}{N_0}\right)\right)$ is represented as a function of log$(\sigma_f)$, for 50 different realizations of the model at a given strain-rate ($\dot{\varepsilon}=10^{-15}$) and temperature ($\theta=8.28\times10^{-3}$) but a varying size $N_0$. If the data for a given size roughly align on such a Weibull's plot, the empirical slope $m$ increases with increasing $N_0$, like for experimental observations \cite{naito2023effect}, and the results obtained at different $N_0$ clearly do not collapse. This invalidates the weakest-link scenario at the bundle scale \cite{weiss2014finite}.

The fact that the bundle strength is not dictated by the strength of its weakest fiber is not surprising, as the mechanical behavior of the bundle results from the cumulative effect of successive fiber ruptures. The simple scenario leading to eqs. (\ref{eq:SScurve}) and (\ref{eq:failureStrength}), which considers such cumulative effect, does not predict any size effect. On the other hand, the bundle rupture, for a classical athermal FBM with equal-load-sharing and under a \textit{stress-control} mode, can be interpreted as a critical phase transition with the failure stress identified as the critical point \cite{pradhan2010failure}. Consequently, finite-size effects are expected for both the mean strength $\langle\sigma_f\rangle$ and the standard deviation $\delta\sigma_f$ \cite{weiss2014finite,zapperi2012current}:

\begin{equation}
    \langle\sigma_{f}\rangle(N_0)=AN_0^{-1/\nu}+\sigma_{\infty} 
    \text{ and  } \delta\sigma_{f}(N_0)=BN_0^{-1/\nu}
    \label{eq:finite-size-effect}
\end{equation}

, where $\nu$ is a finite-size exponent, $\sigma_\infty$ a non-vanishing asymptotic strength for $N_0\rightarrow\infty$, and $A$ and $B$ constants. Therefore, unlike the prediction of a weakest-fiber framework (see above), the average strength of a very large bundle does not vanish, but its associated variability does.
For a uniform distribution of fiber strengths under this stress-control mode and without thermal activation, it has been shown that $\nu=1.5$ and $\sigma_\infty=0.25$ \cite{roy2013scaling}, but both $\nu$ and $\sigma_\infty$ may depend on disorder, i.e. on the distribution of fiber strengths \cite{roy2015fiber}. These finite-size effects are associated to growing avalanche sizes as approaching the critical point at $\sigma_f$. Such framework has been proved relevant to explain experimental size-effects on the compressive failure of disordered materials such as concrete or rocks \cite{weiss2014finite,vu2018revisiting,vu2019compressive}.  

\begin{figure}
\includegraphics[height=0.8\textheight, keepaspectratio]{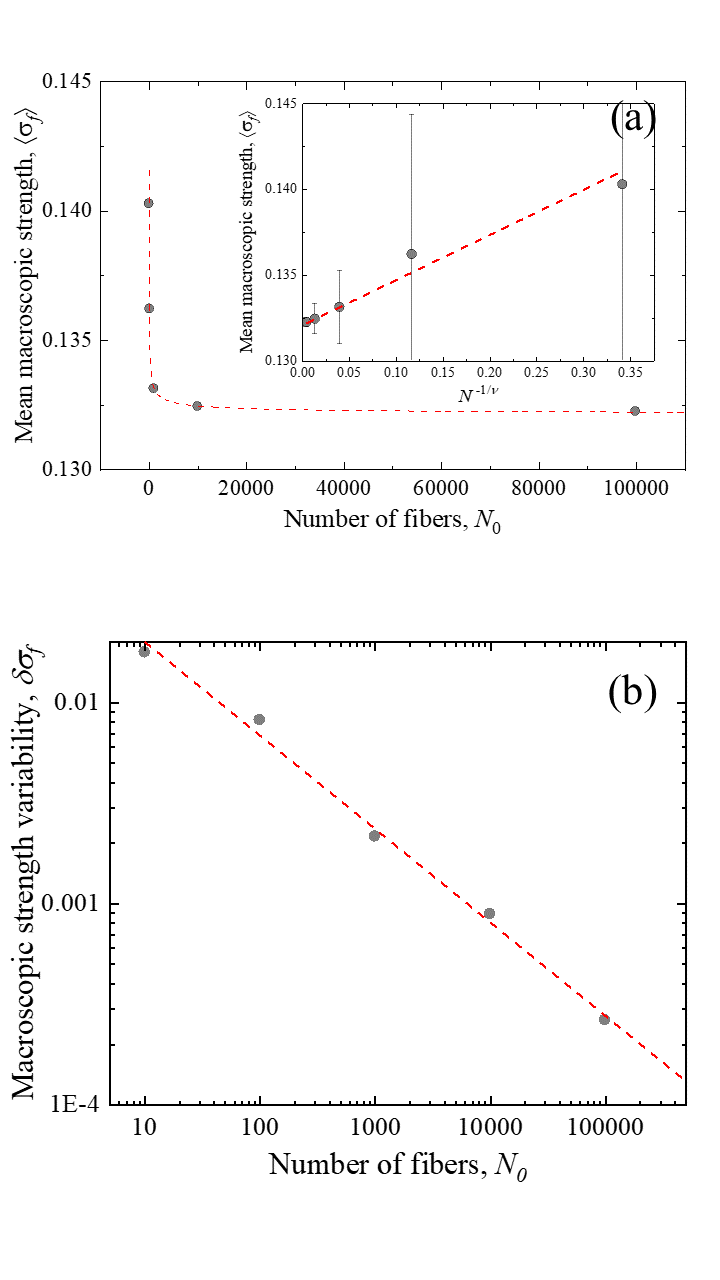}
\caption{\label{fig:Size_effects} Evolution of the mean macroscopic bundle strength $\langle\sigma_f\rangle$ (a) as well as of the associated variability $\delta\sigma_f$ (b) as a function of the initial number of fibers $N_0$ in the bundle, for a Weibull's distribution of fiber strengths with $m=8$, a strain-rate $\dot{\varepsilon}=10^{-15}$ and a temperature $\theta=8.28\times10^{-3}$. The red dashed lines represent eqs. (\ref{eq:finite-size-effect}) with $\nu=2.15$, $\sigma_\infty=0.132$, $A=0.0263$ and $B=0.0577$. On figure (a), the data are represented in a linear scale on the main graph, and on a rescaled graph in the inset. In this inset, the error bars represent the variability values shown on panel (b).}
\end{figure}

The model we consider here fundamentally differs from classical FBM on two points, (i) the introduction of thermal activation and (ii) the \textit{strain-rate-control} mode. It has been already noted that the strain-rate control mode precludes the triggering of athermal avalanches. Still, the finite-size scaling of eq. (\ref{eq:finite-size-effect}) nicely fits the numerical results of fig. (\ref{fig:Size_effects}), with $\nu=2.15$ for $\dot{\varepsilon}=10^{-15}$ and $T=8.28\times10^{-3}$. In this example, the size effect on the mean strength $\langle\sigma_f\rangle$ is moderate, but much more pronounced on the variability $\delta\sigma_f$. The reason why such finite-size scaling still holds in the present case, in the absence of athermal avalanches of fiber breakings, remains to be understood, and it out of the scope of the present paper. In any case, this predicts a non-vanishing mean strength for $N_0\rightarrow\infty$.

\section{\label{sec:Discussion}Discussion}

\subsection{\label{sec:Numbers}Dimensional units}
The results were presented above in terms of dimensionless variables, which are easier to handle and independent of a specific material. Of course, all the trends and conclusions described above would not change if dealing with real numbers. In order to make a more quantitative link to real materials, we can consider Kevlar aramid fibers. Their tensile strength is $\sigma_{f,Kevlar}=3\cdot10^9 Pa$ \cite{Kevlar}, while the scale parameter of our Weibull distributions is $\sigma_u=0.5$ (dimensionless units). Performing creep tests at different temperatures on Kevlar fibers, Giannopoulos and Burgoyne \cite{giannopoulos2012accelerated} reported associated activation volumes $V_a\simeq0.1 nm^3$. From our definition of the stress scale $\theta$ and relation (\ref{eq:rates}), one can write:
\begin{equation}
\frac{\sigma_u}{\theta}=\frac{V_a\cdot\sigma_{f,Kevlar}}{k_BT} \Rightarrow T=\frac{V_a\cdot\theta\cdot\sigma_{f,Kevlar}}{k_B\cdot\sigma_u}
\label{eq:KevlarT}
\end{equation}
With the values reported above and $k_B=1.38\times10^{-23}J\cdot K^{-1}$, one gets $T=4.35\times10^4\times\theta$. Consequently, the 'vanishing' temperature considered on fig. \ref{fig:Athermal_uniform} would correspond to $T=1.2 K$, and the intermediate temperature considered in section \ref{subsec:strain-rate effects} to $T=360 K$. On fig. \ref{fig:temperature_effects}, the temperature corresponding to an almost vanishing bundle strength ($\theta=1.66\times10^{-2}$) would be $T=721K$, i.e. a value close to the crystalline melting point of Kevlar, $\sim810-835K$ \cite{Brown1977Thermal}. We recall here that these numbers are just given as a rough illustration, and for a specific material, Kevlar.

Finally, the attempt frequency $\omega_0=10^{13}s^{-1}$ for rupture processes is generally considered to be of
the same order of magnitude for different solids and independent of the structure and chemical nature of the solid \cite{zhurkov1965kinetic}. With this reference, the strain-rate considered in section \ref{subsec:temperature effects} would correspond to $\dot{\varepsilon}=10^{-2}s^{-1}$, while we explored in section \ref{subsec:strain-rate effects} a range corresponding to $10^{-8}s^{-1}\leq\dot{\varepsilon}\leq 10^{12}s^{-1}$. Of course, the results presented for strain-rates much larger than $1s^{-1}$ are hardly interpretable quantitatively, as in real tests the quasi-static hypothesis would be irrelevant at such very large strain-rates. They were given here only to illustrate the fact that, with this simple model, the athermal mechanical behavior is recovered for strain-rates approaching $\dot{\varepsilon}_{ref}=\omega_0$, as expected.

\subsection{\label{sec:Downscaling}Downscaling from the bundle to the individual fiber scale}

\begin{figure}
\includegraphics[height=0.8\textheight, keepaspectratio]{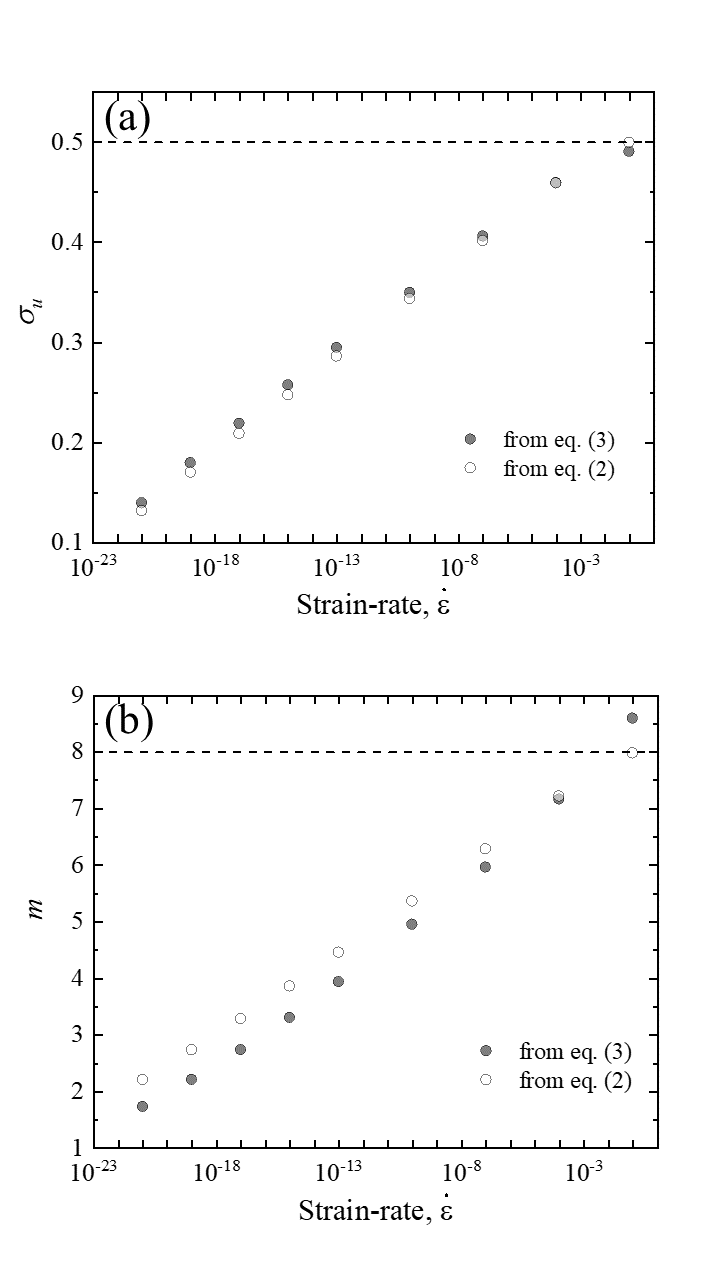}
\caption{\label{fig:Apparent_Weibull_parameters} \textit{Apparent} Weibull's parameters $\sigma_u$ (a) and $m$ (b) of the distribution of individual fiber strengths as estimated from either eq. (\ref{eq:failureStrength}) (gray symbols) or eq. (\ref{eq:SScurve}) (open symbols), for simulations performed with an imposed Weibull's distribution with $\sigma_u=0.5$ and $m=8$, a temperature $\theta=8.28\times 10^{-3}$, and different applied strain-rates. In both panels, the horizontal dashed line represents the imposed, or intrinsic value of the Weibull's parameter. This intrinsic value is only recovered by such analysis for simulations performed under very large strain-rates.}
\end{figure}

As recalled in the introduction, a determination of Weibull's statistics at the fibers scale from a downscaling of the mechanical behavior at the bundle scale has become a common practice \cite{chi1984determination,callaway2017strengths,andersons2002glass,r2008statistical}. As already stressed, this assumes that the mechanical response of the bundle results from a succession of \textit{deterministic} fiber ruptures, from the weakest to the strongest one, ignoring by construction thermally activated ruptures. Applying this methodology to fiber bundle tests performed e.g. at different strain-rates \cite{wang1998effects,wang2024effects,bai2020quasi} can lead to the conclusion that the intrinsic (athermal) distribution of individual fiber strengths depends on strain-rate, as the bundle macroscopic behavior is strain-rate dependent as the result of thermal activation. We already mentioned that individual fibers can creep under constant load, signing the potential role of thermal activation on their mechanical behavior \cite{coleman1957time,wagner1986lifetime}, also illustrated by strain-rate effects on the strength of individual fibers \cite{wang2020tensile}. However, this would not correspond to the intrinsic (athermal) strength considered by the downscaling procedure just mentioned, in particular because thermal activation would introduce, for a given fiber, a strength variability related to thermal disorder. 

To illustrate this problem, one can consider the bundle SS-curves shown on fig. \ref{fig:Strain_rate_effects}(b), corresponding to a Weibull's distribution of individual fiber strengths with $m=8$, to different strain-rates, and $\theta=8.28\times10^{-3}$. From this, the \textit{apparent} Weibull's parameters $m$ and $\sigma_u$ can be obtained, either from the bundle strength $\sigma_f$ and the corresponding strain at failure $\varepsilon_f$, inverting eq. (\ref{eq:failureStrength}), or alternatively from a best-fitting of the SS-curves by eq. (\ref{eq:SScurve}). Doing so, the apparent (stress) scale parameter $\sigma_u$ is observed to grow with the logarithm of the strain-rate (Fig. \ref{fig:Apparent_Weibull_parameters}(a)), and this dependence can be explained qualitatively as in section \ref{subsec:strain-rate effects}. 
The intrinsic (i.e. initially imposed in the simulations) value is recovered only at very large strain-rates when the effect of thermal activation vanishes. The same is true for the shape parameter $m$, see Fig. \ref{fig:Apparent_Weibull_parameters}(b), which decreases as well with decreasing strain-rate. Consequently, the ratio between the standard deviation and the mean of the apparent strength distribution, which reads $\frac{[\Gamma(1+2/m)-(\Gamma(1+1/m))^2]^{1/2}}{\Gamma(1+1/m)}$, is increasing with decreasing strain-rate. This can be interpreted as an increasing relative variability of the fibers strength as the result of thermal disorder.
Finally, we note that, if the two methods (based either on eq. (\ref{eq:failureStrength}) or eq. (\ref{eq:SScurve})) give similar results, the one based on the analysis of the entire bundle SS-curve appears more accurate, in the sense that the Weibull's parameters of the fiber strength distribution are perfectly recovered at large enough strain-rates, which not the case for the simpler method based on eq. (\ref{eq:failureStrength}). This is consistent with \cite{callaway2017strengths}.

This analysis shows that estimating the Weibull's parameters of the fiber strength distribution from the mechanical behavior of a bundle should be done with caution. In particular, this should be performed under experimental conditions ensuring negligible strain-rate and temperature effects on the bundle behavior. Practically, this means running tensile tests on the bundles at large enough strain-rates. However, at very large strain-rates ($\dot{\varepsilon}>1s^{-1}$), dynamical effects not considered in this study may appear as well.

In addition, the bundle behavior may depend on the number of fibers $N_0$, see section \ref{subsec:size effects}, and this could potentially impact this downscaling procedure as well. Here, the recommendation would be to perform mechanical tests on bundles containing a large enough number of fibers to ensure that finite-size corrections are negligible compared to the asymptotic strength $\sigma_\infty$, see eq. (\ref{eq:finite-size-effect}) and fig. \ref{fig:Size_effects}.

\subsection{\label{sec:Depinning}Comparison with the depinning transition of elastic interfaces}

It has been shown above that the macroscopic strength of fiber bundles is characterized, both experimentally and for a thermally-activated democratic FBM, by strain-rate strengthening (Fig. \ref{fig:Strain_rate_effects_2}) and temperature weakening (Fig. \ref{fig:Strength_vs_T}), these two effects being linked through a competition of timescales as described in section (\ref{subsec:strain-rate effects}). At this point, it might be interesting to compare this with the behavior of other driven disordered elastic systems, such as the depinning of elastic interfaces. In the athermal case under a stress- or force-control mode, both the failure of a bundle \cite{pradhan2010failure} or the depinning of an elastic interface \cite{narayan1993threshold} can be interpreted as a critical phase transition and, consequently, characterized by finite-size effects of the type of eq. (\ref{eq:finite-size-effect}). The main difference is of course the behavior beyond the critical point: for an applied force $F>F_c$, the elastic interface moves at an average velocity $v\sim(F-F_c)^\beta$, with $\beta$ the depinning exponent, whereas the bundle is fully broken at $\sigma_M=\sigma_c$. 

The motion of elastic interfaces in a disordered landscape and in the presence of thermal noise has been studied, however essentially under a force-controlled mode. For very low applied forces ($F\ll F_c$) and low temperatures $\theta$, this corresponds to the so-called "creep" regime of depinning \cite{ferrero2021creep}, with the average velocity scaling as $v(F,\theta)\sim exp[-\frac{U_c}{\theta}(\frac{F_c}{F})^\mu]$, where $U_c$ is an energy scale and $\mu$ a characteristic exponent depending on the relative strength of the disorder compared to temperature \cite{kolton2005creep}. From a naive inversion of this, one could expect, under a velocity-controlled mode, a velocity strengthening of the average force, however not in a logarithmic form, as well as a non-linear temperature weakening. However, this regime is hardly comparable to the present work as the condition $F\ll F_c$ means in particular that the interface can locally move back and forth \cite{vandembroucq2004universal}, whereas bundle breaking is by construction an irreversible process. 

The thermally activated regime with an applied force close to the depinning threshold $F_c$, called "thermal rounding", has been studied essentially under a constant force loading mode \cite{vandembroucq2004universal,bustingorry2008thermal,purrello2017creep}. In this case, backward motion can be neglected, meaning that each forward motion becomes "irreversible", and the average velocity exhibits an Arrhenius dependence with a temperature-dependent prefactor, $v(F,\theta)\sim \theta^{-\gamma}exp[-\frac{(F_c-F)}{\theta}]$ \cite{vandembroucq2004universal,purrello2017creep}, for the case $\theta\ll F_c-F$ and small $v$, with $\gamma$ a positive exponent. The Arrhenius dependence can be interpreted as a macroscopic transcription of a microscopic thermal activated mechanism, as described by the equivalent of eq. (\ref{eq:rates}). An inversion of this expression suggests $F_c-\langle F\rangle\sim \theta(-\gamma \text{log}(\theta)-\text{log}(v))$, which implies (i) a temperature weakening of the average force $\langle F\rangle$, linear for small $\theta$ when the logarithmic correction $\text{log}(\theta)$ can be neglected and (ii) a logarithmic velocity strengthening. Recently, the thermally activated motion of a 2D frictional interface, which can be mapped onto the depinning problem in the athermal case \cite{aragon2012seismic}, has been studied under a velocity-controlled mode \cite{weiss2026}, confirming a linear temperature dependence at small $\theta$ and velocity, $F_c-\langle F\rangle\sim \theta$. Interestingly, once translated in terms of a difference between the bundle strength and the corresponding athermal strength, $\sigma_{f,ath}-\sigma_f$, the above expectations (i) and (ii) appear qualitatively consistent with the results shown respectively on fig. \ref{fig:Strain_rate_effects_2} and fig. \ref{fig:Strength_vs_T}, as well as with the competition of timescales presented in section \ref{subsec:strain-rate effects}. This hints at a possible extension of the results presented here to other driven disordered elastic systems, and calls for more detailed studies of activated depinning under a velocity-controlled mode. 

\section{\label{Conclusion}Conclusion}

Tensile tests performed on individual fibers or bundles of various materials revealed strain-rate and temperature effects on strength, strain at peak stress, and apparent Young's modulus. In addition, fibers and bundles can creep under a constant load, all of this arguing for a key role of thermal activation on their mechanical behavior. Here, it was proposed to rationalize these various effects, as well as size (number of fibers in a bundle) effects, from a fiber-bundle model with equal-load-sharing in which thermal activation of fiber breakings is introduced from a kinetic Monte-Carlo algorithm adapted for time-varying stresses. In agreement with experimental data, this model shows:
\begin{itemize}
    \item A logarithmic decrease of the bundle strength $\sigma_f$ as well as of the strain at peak stress $\varepsilon_f$ with decreasing strain-rate.
    \item A decrease of the apparent Young's modulus of the bundle with decreasing strain-rate.
    \item A softening of the bundle, i.e. a decrease of the bundle strength as well as of the strain at peak stress, with increasing temperature.
    \item A decrease of the mean bundle strength $\langle\sigma_f\rangle$, as well as of the associated variability $\delta\sigma_f$ among different realizations of disorder, with an increasing number of fibers in the bundle. These size effects cannot however be explained from a weakest-fiber framework, and the mean strength rapidly saturates for a large enough number of fibers in the bundle.
\end{itemize}

These strain-rate and temperature effects can be explained from a competition of timescales between a timescale set by the inverse of the applied strain-rate, and a typical thermal activation timescale. On the other hand, these strain-rate and temperature effects, as well as the role of thermal activation on fiber breaking and bundle mechanical behavior imply that an estimation of the distribution of individual fiber strengths, and particularly the associated Weibull's parameters $m$ and $\sigma_u$, from a downscaling of the mechanical behavior at the bundle scale, should be considered with caution. If the mechanical testing of the bundle is performed under conditions favouring the role of thermal activation (e.g. low applied strain-rate), this procedure can strongly underestimate the intrinsic (athermal) values of $m$ and $\sigma_u$.

\section*{Acknowledgments}
This work has been partly supported by the French National Research Agency, project DISCREEP (ANR-23-CE30-0031-01)
\\

\bibliography{BiblioFiberBundles}

\end{document}